\documentclass[twocolumn]{aastex62}
\usepackage{hyperref}
\usepackage{amsmath,amstext}
\usepackage[T1]{fontenc}
\usepackage[figure,figure*]{hypcap}
\usepackage{graphicx}
\usepackage{multirow}

\def\sun{$_{\odot}$}

\def\f28{${f}_{2-8{\rm keV}}$}

\def\sun{$_{\odot}$}

\def\ergscm2{erg s$^{-1}$ cm$^{-2}$}
\def\yr-1{yr$^{-1}$}

\def\asec{\ifmmode^{\prime\prime}\else$^{\prime\prime}$\fi}

\received{September 7, 2019}
\submitjournal{ApJ}

\shorttitle{Tracing the AGN/X-ray Binary Analogy with Individual Changing-Look AGN} 
\shortauthors{Ruan {\it et al.}}

\begin{document}

\title{Tracing the AGN/X-ray Binary Analogy with Light Curves of Individual Changing-Look AGN}

\correspondingauthor{John~J.~Ruan}
\email{john.ruan@mcgill.ca}

\author[0000-0001-8665-5523]{John~J.~Ruan}
\affil{McGill Space Institute and Department of Physics, McGill University, 3600 rue University, Montreal, Quebec, H3A 2T8, Canada}

\author{Scott~F.~Anderson}
\affil{Department of Astronomy, University of Washington, Box 351580, Seattle, WA 98195, USA}

\author{Michael~Eracleous}
\affil{Department of Astronomy \& Astrophysics and Institute for Gravitation and the Cosmos, The Pennsylvania State University, 525 Davey Lab, University Park, PA 16802, USA}

\author{Paul~J.~Green}
\affil{Harvard Smithsonian Center for Astrophysics, 60 Garden St, Cambridge, MA 02138, USA}

\author[0000-0001-6803-2138]{Daryl Haggard}
\affil{McGill Space Institute and Department of Physics, McGill University, 3600 rue University, Montreal, Quebec, H3A 2T8, Canada}
\affil{CIFAR Azrieli Global Scholar, Gravity \& the Extreme Universe Program, Canadian Institute for Advanced Research, 661 University Avenue,
Suite 505, Toronto, ON M5G 1M1, Canada}

\author{Chelsea~L.~MacLeod}
\affil{Harvard Smithsonian Center for Astrophysics, 60 Garden St, Cambridge, MA 02138, USA}

\author{Jessie~C.~Runnoe}
\affil{Department of Astronomy, University of Michigan, 1085 S. University Avenue, Ann Arbor, MI 48109, USA}

\author{Malgosia~A.~Sobolewska}
\affil{Harvard Smithsonian Center for Astrophysics, 60 Garden St, Cambridge, MA 02138, USA}

\begin{abstract}
Physical models of X-ray binary outbursts can aid in understanding the origin of `changing-look' active galactic nuclei (AGN), if we can establish that these two black hole accretion phenomena are analogous. Previously, studies of the correlation between the UV-to-X-ray spectral index $\alpha_\mathrm{OX}$ and Eddington ratio using single-epoch observations of changing-look AGN \emph{samples} have revealed possible similarities to the spectral evolution of outbursting X-ray binaries. However, direct comparisons using multi-epoch UV/X-ray light curves of \emph{individual} changing-look AGN undergoing dramatic changes in Eddington ratio have been scarce. Here, we use published \emph{Swift} UV/X-ray light curves of two changing-look AGN (NGC~2617 and ZTF18aajupnt) to examine the evolution of their $\alpha_\mathrm{OX}$ values during outburst. We show that the combination of these two changing-look AGN can trace out the predicted spectral evolution from X-ray binary outbursts, including the inversion in the evolution of $\alpha_\mathrm{OX}$ as a function of Eddington ratio. We suggest that the spectral softening that is observed to occur below a critical Eddington ratio in both AGN and X-ray binaries is due to reprocessing of Comptonized X-ray emission by the accretion disk, based on the X-ray to UV reverberation lags previously observed in NGC~2617. Our results suggest that the physical processes causing the changing-look AGN phenomenon are similar to those in X-ray binary outbursts.
\end{abstract}

% ============================
\section{Introduction}
\label{sec:intro}

	Although active galactic nuclei (AGN) are famously variable at nearly all wavelengths \citep[e.g.,][]{ulrich97}, discoveries of new and rare AGN variability phenomena have been rapidly increasing. Detection of unusual AGN variability using light curves from wide-field time-domain imaging surveys has enabled rapid multi-wavelength follow-up observations, which can characterize and unveil the origin of these phenomena. Amongst these discoveries are many dozens of `changing-look AGN', which display a characteristic sudden appearance or disappearance of their optical broad emission lines, often accompanied by dramatic photometric variability (e.g., $>$1 mag in the optical) \citep[e.g.,][]{shappee14, denney14, mcelroy16, parker16, trakhtenbrot19, katebi18}. Similar behavior has also been observed in AGN at higher redshifts and luminosities \citep[e.g.,][]{lamassa15, ruan16, runnoe16, macleod16, gezari17, yang18, wang18, macleod19, graham19a, sheng19, frederick19}, and these objects  have been referred to as  `changing-look quasars'\footnote{Since changing-look AGN and changing-look quasars detected in optical spectroscopy are both likely to be undergoing similar physical changes, we will collectively refer to all these objects as `changing-look AGN' here in this paper.}. Since the broad line gas in AGN is photoionized by the UV continuum, the appearance/disappearance of the broad emission lines occurs in tandem with dramatic variations in the UV/optical continuum, with a time-lag that is consistent with expectations from reverberation mapping studies \citep{trakhtenbrot19}. However, many aspects of the changing-look AGN phenomena remain unexplained. 

	A still-puzzling property of changing-look AGN is the surprisingly short timescales of just a few years for their dramatic transformations. Nearly all observational tests have now disfavored transient dust obscuration or nuclear tidal disruption events as the source of the observed fading/brightening of the broad emission lines and continuum emission \citep{lamassa15, ruan16, runnoe16, macleod16, hutsemekers17, sheng17, yang18, stern18, macleod19, hutsemekers19, dexter19a}. Instead, dramatic changes intrinsic to the accretion flow of changing-look AGN, such as those associated with accretion state transitions commonly observed in outbursting X-ray binaries \citep[e.g.,][]{homan05, remillard06, done07}, can cause the luminosity of the accretion disk to change dramatically. This in turn can cause the broad emission lines to appear or disappear, especially in some models where the broad line region gas is associated with winds from a luminous accretion disk \citep[e.g.,][]{murray95, murray97, elitzur09, elitzur14, elvis17}. Intriguingly, multi-epoch observations of changing-look AGN often show that they cross a critical bolometric Eddington ratio of $L_\mathrm{bol}/L_\mathrm{Edd}$ $\sim$ 10$^{-2}$ during their dramatic transformations \citep{macleod19, ruan19}, approximately where accretion state transitions are often observed to occur in X-ray binaries \citep[e.g.,][]{maccarone03}. This could be consistent with a scenario in which changing-look AGN are undergoing accretion state transitions (e.g., between a luminous thin disk in the high-luminosity/soft-spectrum state and a radiatively inefficient accretion flow in the low-luminosity/hard-spectrum state), analogous to X-ray binary outbursts. However, state transitions in X-ray binaries are observed to occur over timescales of $\sim$days, and a simple scaling of this transition time from stellar mass black holes to supermassive black holes suggests that analogous transitions would occur in AGN over timescales of $\sim$10$^{5}$~years. Due to this large discrepancy between the observed timescales for changing-look AGN and the expected timescales for state transitions in AGN, it is unclear whether the changing-look AGN phenomenon is analogous to X-ray binary outbursts.

	Many theoretical models have been proposed to explain the short timescales for the dramatic transformations in changing-look AGN. \citet{stern18} emphasize that the observed timescales are most consistent with the thermal or heating/cooling front propagation timescale ($\sim$1 yr). Disk truncation, often invoked to interpret accretion state transitions in X-ray binaries, would occur on the much longer viscous timescale ($\sim$400 yr). This argument would be consistent with the model detailed in \citet{ross18}, in which the UV continuum in luminous changing-look AGN can change dramatically due to a change in torques across the inner disk, which then causes a cooling/heating front to propagate radially into the outer disk. \citet{sniegowska19} instead suggest that for changing-look AGN at $L_\mathrm{bol}/L_\mathrm{Edd}$ $\sim$ 10$^{-2}$, a radiation pressure instability at the boundary between the inner advection dominated accretion flow \citep[ADAF;][]{narayan94} and the truncated thin disk can cause the accretion rate and luminosity to vary dramatically. \citet{dexter19b} propose that variability timescales in all quasars can decrease if their accretion disks are geometrically thick, which may occur in the case of strong magnetic pressure support. \citet{noda18} argue that at least some changing-look AGN are indeed undergoing state transitions similar to X-ray binaries, and strong radiation and/or magnetic pressure in the disk shortens the transition timescales to the observed timescales of a few years. Our results here are consistent with this state transition interpretation, based on comparing the observed UV-to-X-ray spectral energy distribution (SED) changes in changing-look AGN to X-ray binary outbursts.

	Since the hallmark of accretion state transitions in X-ray binaries is the characteristic evolution of their SED as a function of luminosity, we can test whether changing-look AGN are analogous to X-ray binary outbursts by comparing the evolution of their thin accretion disk and Comptonized coronal emission as a function of Eddington ratio. In X-ray binaries, the thin disk emission peaks in the soft X-rays, while the Comptonized emission dominates the hard X-rays. The evolution of their X-ray hardness is thus a probe of the geometry of their disk-corona systems during state transitions. For AGN, the thin disk emission is thought to peak in the UV, while the Comptonized emission dominates the X-rays. Thus, the UV-to-X-ray spectral index \citep[$\alpha_\mathrm{OX}$; ][]{tananbaum79} probes the disk-corona geometry in AGN, and is analogous to the X-ray hardness in X-ray binaries. In this way, comparisons of the observed evolution of $\alpha_\mathrm{OX}$ in changing-look AGN to the evolution of the X-ray hardness in outbursting X-ray binaries can reveal whether the accretion flows in these two phenomena are undergoing analogous changes.

	Previously, single-epoch UV and X-ray observations of \emph{samples} of AGN (including changing-look AGN) have suggested that they may display similar spectral evolution as X-ray binaries. For X-ray binaries fading from outburst, their X-ray spectra are observed to harden as their bolometric Eddington ratios decrease from near-Eddington to $L_\mathrm{bol}/L_\mathrm{Edd}$ $\sim$ 10$^{-2}$. This spectral hardening has been interpreted as resulting from the progressive evaporation of the inner disk \citep{esin97}, resulting in an inner ADAF that is surrounded by a truncated thin disk. As the X-ray binary further fades below Eddington ratios of $L_\mathrm{bol}/L_\mathrm{Edd}$ $\lesssim$ 10$^{-2}$, the X-ray spectra are often observed to soften \citep[e.g.,][]{ebisawa94, revnivtsev00, tomsick01, corbel04, kalemci05, wu08, russell10, homan13, kalemci13, kajava16, plotkin17}, creating a characteristic `V-shape' inversion in the evolution of the X-ray hardness as a function of $L_\mathrm{bol}/L_\mathrm{Edd}$ \citep[see e.g., Figure~1 of ][]{sobolewska11a}. If changing-look AGN are undergoing accretion state transitions analogous to X-ray binary outbursts, they would be expected to follow a similar V-shape inversion in the evolution of their $\alpha_\mathrm{OX}$ values as a function of $L_\mathrm{bol}/L_\mathrm{Edd}$. For samples of AGN above $L_\mathrm{bol}/L_\mathrm{Edd}$ $\gtrsim$ 10$^{-2}$, previous single-epoch UV/X-ray observations have consistently revealed a positive correlation between $\alpha_\mathrm{OX}$ and $L_\mathrm{bol}/L_\mathrm{Edd}$ \citep[e.g.,][]{vignali03, strateva05, steffen06, just07, grupe10, jin12, wu12, vagnetti13, trichas13}. This positive correlation implies a hardening of the SED as $L_\mathrm{bol}/L_\mathrm{Edd}$ decreases towards $\sim$10$^{-2}$, and is consistent with the right side of the V-shape evolution of $\alpha_\mathrm{OX}$ that is expected from X-ray binaries. However, the left side of this V-shape evolution, in which the SED softens below $L_\mathrm{bol}/L_\mathrm{Edd}$ $\lesssim$ 10$^{-2}$, has been difficult to observe in AGN due to a variety of issues. \citet{ruan19} used single-epoch UV/X-ray observations of a sample of faded changing-look AGN with current $L_\mathrm{bol}/L_\mathrm{Edd} \sim 10^{-2}$ to $\sim$10$^{-3.5}$ to show that their $\alpha_\mathrm{OX}$ is anti-correlated with $L_\mathrm{bol}/L_\mathrm{Edd}$. This negative correlation implies a softening of the SED below $L_\mathrm{bol}/L_\mathrm{Edd}$ $\lesssim$ 10$^{-2}$, consistent with the left side of the V-shape evolution of $\alpha_\mathrm{OX}$ expected from X-ray binaries (see their Figure~\ref{fig:vshape}). That work not only suggests that the disk-corona geometry of changing-look AGN is analogous to X-ray binary outbursts, but more generally extends the AGN/X-ray binary analogy to $L_\mathrm{bol}/L_\mathrm{Edd}$ $\lesssim$ 10$^{-2}$.

	Ideally, this comparison between the spectral evolution of changing-look AGN and X-ray binary outbursts would be performed using multi-epoch UV and X-ray light curves of \emph{individual} changing-look AGN as they undergo large changes in $L_\mathrm{bol}/L_\mathrm{Edd}$, rather than previous studies that relied on single-epoch observations of AGN \emph{samples} that span a wide range in $L_\mathrm{bol}/L_\mathrm{Edd}$. Using light curves of individual AGN avoids complications inherent to AGN samples, such as the additional effects of differences in black hole mass, dust obscuration, and inclination between the objects in a sample. However, well-sampled UV/X-ray light curves of changing-look AGN are scarce, so the predicted V-shape evolution of their $\alpha_\mathrm{OX}$ values has not been unambiguously demonstrated in an individual AGN. The most intriguing results so far are based on the fading of the changing-look AGN Mrk~1018 \citep{mcelroy16}. \citet{husemann16} use multi-epoch \emph{XMM-Newton} and \emph{Swift} optical/UV/X-ray observations during the fading of this AGN to show that the fading is stronger in the UV than in the X-ray. Based on an independent analysis of these data, \citet{noda18} point out that this result implies a hardening of the SED as the $L_\mathrm{bol}/L_\mathrm{Edd}$ decreases towards $\sim$10$^{-2}$, which is consistent with the AGN moving down the right side of the V-shape inversion that is associated with accretion state transitions in X-ray binaries. However, no  light curves of individual AGN have yet revealed the predicted SED softening below $L_\mathrm{bol}/L_\mathrm{Edd}$ $\lesssim$ 10$^{-2}$ (i.e., moving along the left side of the V-shape inversion).
	
 %------- FIGURE 1 -------
\begin{figure} [t!]
\center{
\includegraphics[scale=0.54,angle=0]{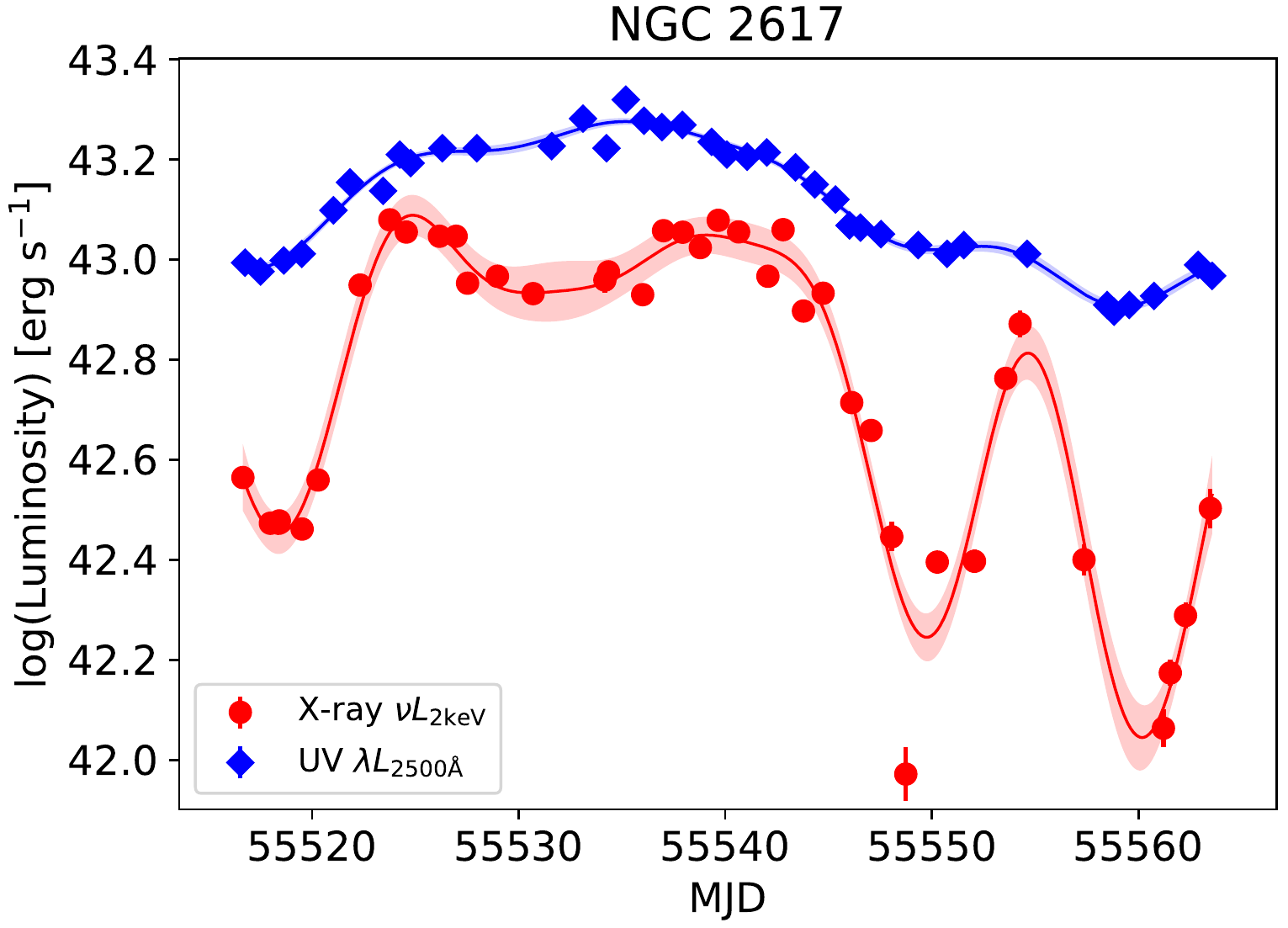}
}
\figcaption{X-ray and UV light curves of the outburst of NGC~2617, as observed by \emph{Swift} and based on fluxes reported by \citet{shappee14}. The 2500~\AA~(blue points) and 2 keV (red points) luminosities are shown.  Gaussian process fits to the two light curves are also shown (red and blue lines), along with their 1$\sigma$ uncertainties (red and blue shaded regions). During the initial brightening of the outburst in both the UV and X-rays (between MJD of approximately 55520 and 55530), the X-ray emission brightens by a larger factor than the UV. This is indicative of a hardening of the UV-to-X-ray spectral index $\alpha_\mathrm{OX}$ as the UV Eddington ratio increases. A similar trend is seen during the fading (between MJD of approximately 55540 and 55550), where the X-ray emission fades by a larger factor than the UV, indicative of a softening of $\alpha_\mathrm{OX}$ as the UV Eddington ratio decreases. 
}
\label{fig:ngc2617_lc}
\end{figure}
% ============================

	Here, we use published Neil Gehrels \emph{Swift} Observatory \citep{gehrels04} UV/X-ray light curves of two changing-look AGN (NGC~2617 and ZTF18aajupnt) undergoing dramatic transformations to investigate their $\alpha_\mathrm{OX}$ evolution. The original investigations that obtained these \emph{Swift} observations yielded several interesting science results, but they did not probe the $\alpha_\mathrm{OX}$ evolution as a function of Eddington ratio in detail. The outline of this paper is as follows: In Section~2, we describe the published \emph{Swift} UV and X-ray observations of the two changing-look AGN during outburst, and our further analysis of the light curves. In Section~3, we present the evolution of their $\alpha_\mathrm{OX}$ as a function of Eddington ratio, compare these observations to X-ray binary outbursts, and describe a reprocessing-based interpretation of the results. We briefly conclude in Section~4. Throughout this work, we assume a standard $\Lambda$CDM cosmology with $\Omega_\mathrm{m} = 0.309$, $\Omega_\Lambda = 0.691$, and $H_0 = 67.7$ km s$^{-1}$ Mpc$^{-1}$ \citep{bennett14}.

\section{Archival \emph{Swift} Observations of two \\ outbursting changing-look AGN}
\label{sec:obs}

\subsection{Light curves of NGC~2617}
\label{ssc:ngc2617}

	NGC~2617 is a nearby Seyfert 2 galaxy at $z = 0.0142$ that underwent an outburst in 2012, before slowly fading. \citet{shappee14} obtained well-sampled multi-epoch \emph{Swift} X-ray and UV/optical observations throughout the outburst, producing light curves that span the initial rise and the subsequent decay (see Figure~\ref{fig:ngc2617_lc}). The main focus of the investigation by \citet{shappee14} was the robust detection of time-lags between the X-ray, UV, and optical light curves. They found that the X-ray emission leads the UV/optical by a time-lag of a few days, suggestive of a model in which the X-ray emission from a central source is reprocessed by the accretion disk to produce the UV/optical emission. Here, we use these published observations of NGC~2617 to instead investigate its SED evolution (based on $\alpha_\mathrm{OX}$) as a function of Eddington ratio throughout its outburst.
	
%------- FIGURE 2 -------
\begin{figure} [t!]
\center{
\includegraphics[scale=0.54,angle=0]{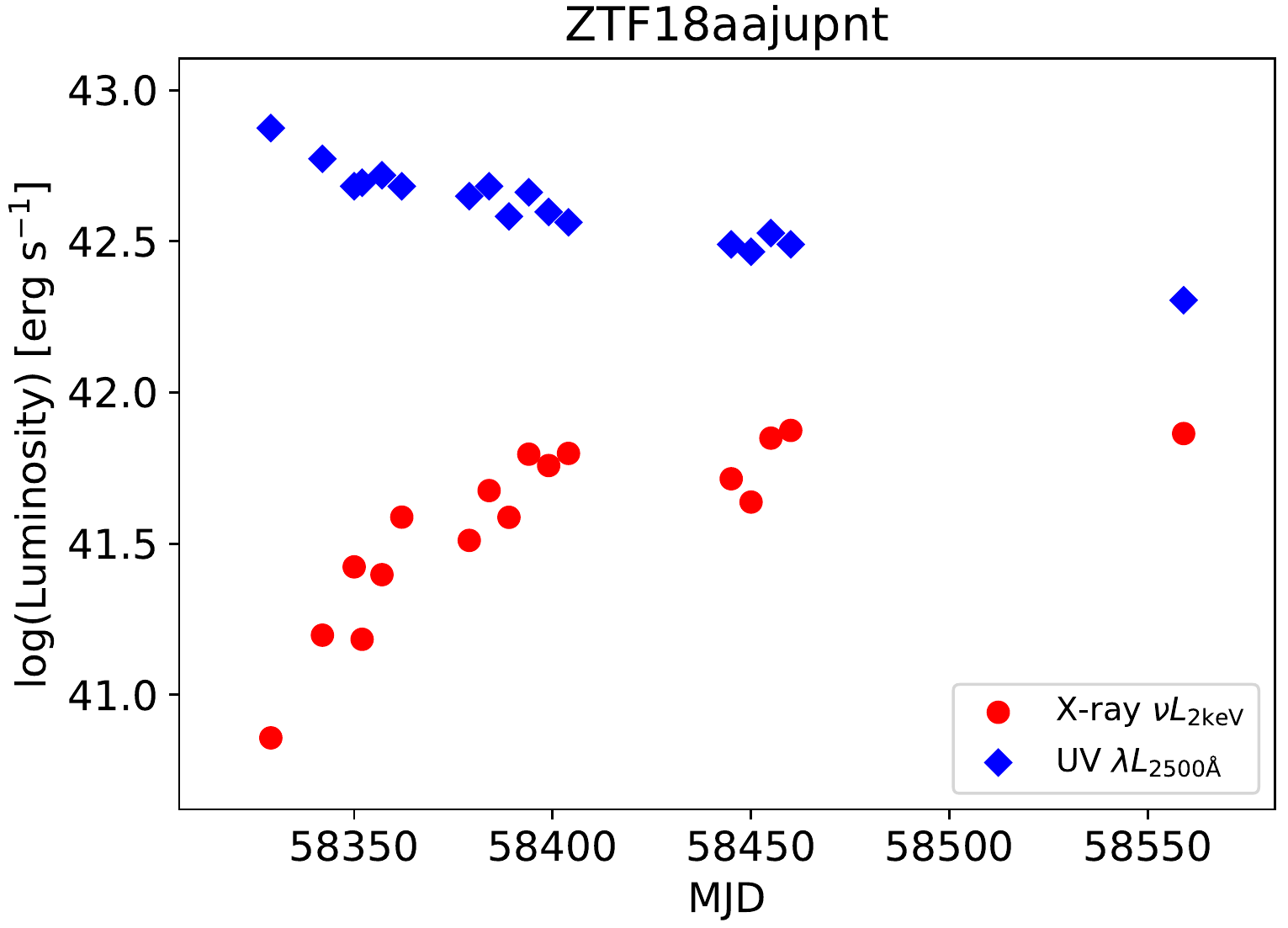}
}
\figcaption{X-ray and UV light curve of ZTF18aajupnt  during its outburst, as observed by \emph{Swift} and based on fluxes reported by \citet{frederick19}. The 2500~\AA~(blue points) and 2 keV (red points) luminosities are shown. While the UV disk emission is fading from its peak, the Comptonized X-ray emission brightens, indicative of a hardening in its SED as the UV Eddington ratio decreases.
}
\label{fig:ztf18aajupnt _lc}
\end{figure}
% ============================

	We use the contemporaneous \emph{Swift} XRT \citep{burrows05} and UVOT \citep{roming05} light curves obtained and reduced by \citet{shappee14} to calculate $\alpha_\mathrm{OX}$, which is defined as 
\begin{equation}
	\alpha_\mathrm{OX} = - \frac{\mathrm{log}(\lambda L_\mathrm{2500\text{\normalfont\AA}}) - \mathrm{log}(\nu L_\mathrm{2keV})}{\mathrm{log}(\nu_\mathrm{2500\text{\normalfont\AA}}) - \mathrm{log}(\nu_\mathrm{2keV})} +1.
 \end{equation}
 To calculate the 2~keV X-ray luminosity ($\nu L_\mathrm{2keV}$) at each epoch of observations during the outburst, we use the unabsorbed $F(\mathrm{0.3-10keV})$ flux and X-ray photon index $\Gamma$ at each epoch reported by \citet{shappee14}, assuming the power-law spectral model that they fitted to the X-ray spectrum. We estimate the 2500~\AA~UV luminosity ($\lambda L_{2500\text{\normalfont\AA}}$) at each epoch by converting the Vega magnitudes in the UVW1 filter, using a photometric zero-point of 17.44 mag \citep{breeveld11} and a count rate to flux density conversion factor of $4.3 \times 10^{-16}$ \citep[for flux density in units of erg s$^{-1}$ cm$^{-2}$ \AA$^{-1}$;][]{poole08}. This UVW1 filter has a central wavelength of 2600~\AA~and full-width at half-max of $\sim$683\AA~\citep{poole08}, thus providing a relatively good estimate of $\lambda L_{2500\text{\normalfont\AA}}$. Since \citet{shappee14} detected a time-lag of 3.22 days between the X-ray and UVW1 light curves in these observations, we add 3.22 days to the observation dates of the X-ray light curve. Although this shift essentially aligns the variability in the X-ray and UVW1 light curves, the individual X-ray and UV epochs in the light curve are no longer contemporaneous. We thus interpolate the shifted X-ray light curve at each epoch of UV observations to produce pairs of $\nu L_\mathrm{2keV}$ and $\lambda L_{2500\text{\normalfont\AA}}$ values to use in Equation~1. To perform this interpolation, we fit the shifted $\nu L_\mathrm{2keV}$ X-ray light curve using a Gaussian process model, and interpolate it at each MJD of the $\lambda L_{2500\text{\normalfont\AA}}$ UV light curve. We perform this Gaussian process modeling using the \texttt{George} software \citep{ambikasaran15}, adopting an exponential-squared covariance function and optimizing the hyperparameters therein. We then remove epochs at the beginning of the UV light curve which have MJD before the first epoch of the shifted X-ray light curve, and we similarly remove epochs at the end of the shifted X-ray light curve which have MJD after the last epoch of the UV light curve, since calculation of $\alpha_\mathrm{OX}$ at these epochs requires extrapolation of either the UV or shifted X-ray light curves, and thus leads to poor constraints on $\alpha_\mathrm{OX}$. Finally, we subtract the host galaxy starlight contribution to the $\lambda L_{2500\text{\normalfont\AA}}$ luminosities. Since no UV observations of NGC~2617 are available prior to its 2012 outburst and more recent observations show that fainter nuclear activity still persists \citep{oknyansky17}, the host galaxy luminosity must be estimated through SED modeling. \citet{shappee14} fitted a model to their multi-band data that incorporates accretion disk and host galaxy starlight components, and showed that their best-fitting host galaxy component (see their Figure~11) is well-described by the Sbc galaxy SED template from \citet{assef10}. We assume the best-fit host galaxy luminosity of $10^{42}$~erg~s$^{-1}$ in the UVW1 filter from this modeling, and subtract this contribution from the $\lambda L_{2500\text{\normalfont\AA}}$ light curve. We examine the effects of this assumption in Section~\ref{ssc:host} in the Appendix, and conclude that our results are not strongly affected by the host galaxy subtraction. 

	The resulting $\lambda L_{2500\text{\normalfont\AA}}$ and shifted $\nu L_\mathrm{2keV}$ light curves of NGC~2617 are shown in Figure~\ref{fig:ngc2617_lc}, along with the fitted Gaussian process models. The $\nu L_\mathrm{2keV}$ light curve has been shifted to account for the time-lag, while the estimated host galaxy contribution has been subtracted from the $\lambda L_{2500\text{\normalfont\AA}}$ light curve, and both light curves have been clipped at the beginning or end, as described above. The $\alpha_\mathrm{OX}$ values are based on the measured $\lambda L_{2500\text{\normalfont\AA}}$ at each epoch, and a $\nu L_\mathrm{2keV}$ interpolated from the Gaussian process fitted to the $\nu L_\mathrm{2keV}$ light curve. Our computed $\lambda L_{2500\text{\normalfont\AA}}$ (with host galaxy subtraction) and interpolated $\nu L_\mathrm{2keV}$ values are listed in Table~1, along with the corresponding $\alpha_\mathrm{OX}$ values.

\subsection{Light curves of ZTF18aajupnt }
\label{ssc:ztf18aajupnt }

	ZTF18aajupnt  is a nearby low-ionization nuclear emission line region (LINER) galaxy at $z = 0.0367$ that underwent an outburst in 2017, transforming into a Type 1 broad line AGN. After the discovery of this outburst in optical imaging from the Zwicky Transient Facility \citep[ZTF;][]{graham19b, bellm19}, \citet{frederick19} obtained well-sampled multi-epoch \emph{Swift} X-ray and UV/optical observations during its decay in the optical. The \emph{Swift} UV light curve shows a slow fading over several months, while the X-ray light curve shows a contemporaneous brightening. This behavior is likely probing a UV-to-X-ray SED evolution that may reflect a change in its disk-corona geometry as the Eddington ratio changes.

	We calculate $\nu L_\mathrm{2keV}$ and $\lambda L_{2500\text{\normalfont\AA}}$ luminosities at each light curve epoch of ZTF18aajupnt. For $\nu L_\mathrm{2keV}$, we use the unabsorbed $F(\mathrm{0.3-10keV})$ fluxes reported by \citet{frederick19}, and assume the power-law spectral model with photon index $\Gamma = 2.82$ measured using the coadded XRT spectrum by \citet{frederick19}. For $\lambda L_{2500\text{\normalfont\AA}}$, we use the values reported by \citet{frederick19}. To estimate the host galaxy contribution to the $\lambda L_{2500\text{\normalfont\AA}}$ luminosities, we use the \emph{GALEX} AB magnitude of $NUV = 19$ mag measured prior to the outburst by \citet{frederick19}. We assume that the host galaxy SED is described by the Sbc galaxy template from \citet{assef10}, similar to the results from SED modeling of NGC~2617 by \citet{shappee14}. We then multiply this template SED with the \emph{GALEX} $NUV$ filter transmission curve, and find that a host galaxy luminosity of $10^{42.44}$~erg~s$^{-1}$ at 2500~\AA~ is required to produce the observed $NUV = 19$ mag. We thus subtract this host galaxy luminosity from our $\lambda L_{2500\text{\normalfont\AA}}$ values, although we show in Section~\ref{ssc:host} in the Appendix that our conclusions are not strongly affected by the host galaxy subtraction. The resulting $\lambda L_{2500\text{\normalfont\AA}}$ (with host galaxy subtraction) and $\nu L_\mathrm{2keV}$ light curves of ZTF18aajupnt are shown in Figure~\ref{fig:ztf18aajupnt _lc}, and their luminosity values are listed in Table~2 along with their corresponding $\alpha_\mathrm{OX}$ values.

 %------- FIGURE 3 -------
\begin{figure*} [t!]
\center{
\includegraphics[scale=0.63,angle=0]{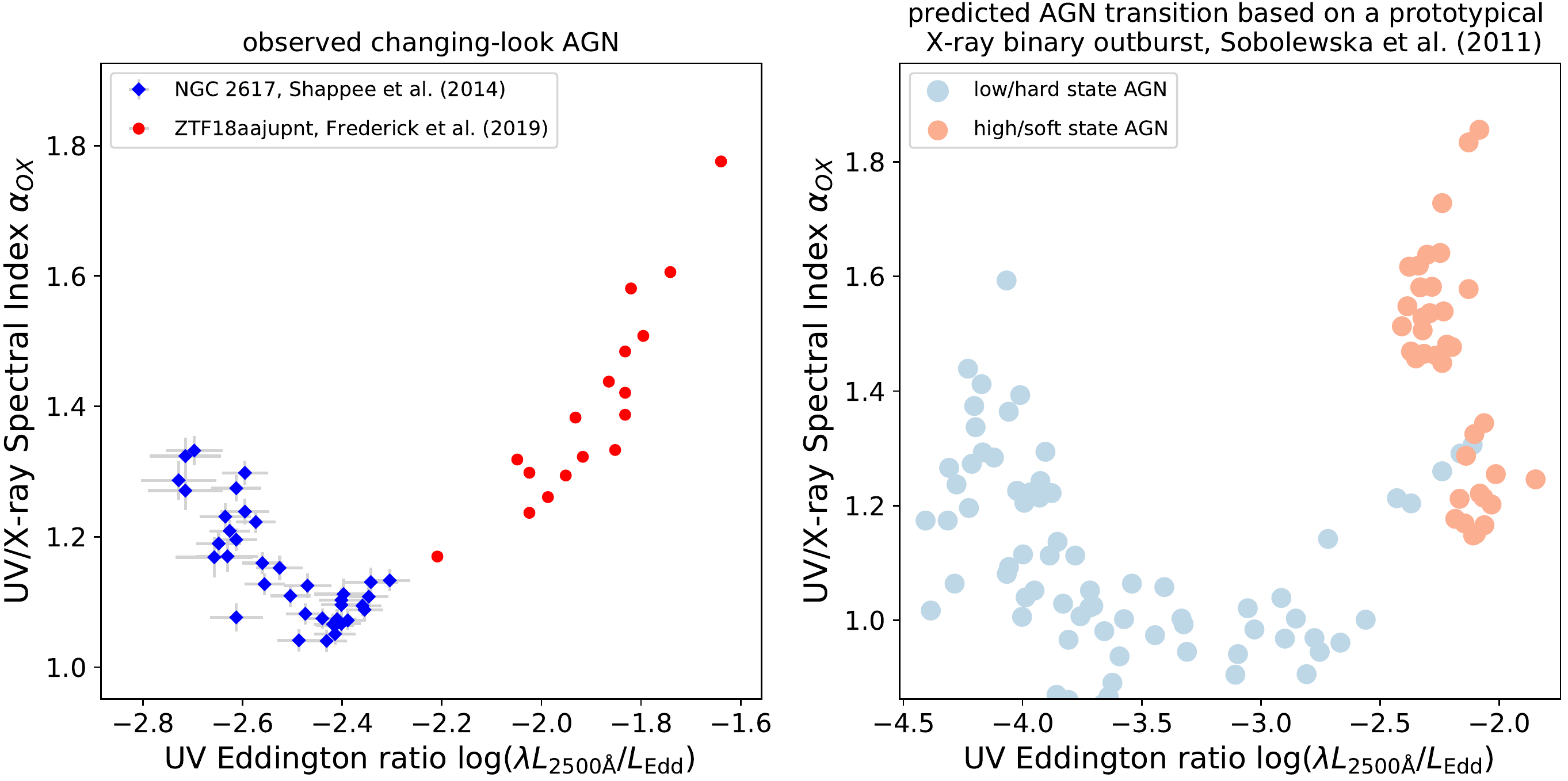}
}
\figcaption{The combined UV-to-X-ray spectral evolution of ZTF18aajupnt and NGC~2617 during their outburst traces out a V-shaped inversion that is predicted from X-ray binary outbursts. \emph{Left}: The SED of  ZTF18aajupnt (red points) hardens as the UV Eddington ratio deceases, leading to a decrease in the UV-to-X-ray spectral index $\alpha_\mathrm{OX}$. At lower UV Eddington ratios, the SED of NGC~2617 (blue diamonds) softens as the UV Eddington ratio deceases, leading to an increase in $\alpha_\mathrm{OX}$. When combined, these two changing-look AGN reveal a clear V-shape inversion in the evolution of their $\alpha_\mathrm{OX}$ values as a function of UV Eddington ratio. \emph{Right}: Prediction for the $\alpha_\mathrm{OX}$ evolution in AGN, based on a prototypical outburst of the X-ray binary GRO~J1655$-$40. This illustrates the predicted V-shape inversion in the evolution of $\alpha_\mathrm{OX}$ in AGN, and is qualitatively similar to the observations in the left panel, thus suggesting that the changing-look AGN phenomenon is analogous to X-ray binary outbursts. In Section~\ref{ssc:repro}, we suggest a qualitative interpretation for this V-shape evolution of $\alpha_\mathrm{OX}$ in which AGN first undergo an accretion state transition from the high/soft state to the low/hard state as their Eddington ratios decrease. Below a critical Eddington ratio, the SED then softens again due to reprocessing of Comptonized X-rays from the corona into the UV/optical by the disk.
}
\label{fig:vshape}
\end{figure*}
% ============================

\subsection{Calculating Eddington ratios}
\label{sec:eddratios}

	We calculate a UV Eddington ratio using $\lambda L_{2500\text{\normalfont\AA}}$/$L_\mathrm{Edd}$ = $\lambda L_{2500\text{\normalfont\AA}}$/($1.26\times10^{38}$ $M_\mathrm{BH}$), for $M_\mathrm{BH}$ in units of $M_\odot$, and  $\lambda L_{2500\text{\normalfont\AA}}$ in units of erg s$^{-1}$. For NGC~2617, we assume $M_\mathrm{BH} = 10^{7.5\pm0.5}$~$M_\odot$ as estimated from reverberation mapping of its broad H$\beta$ emission \citep{fausnaugh17}, which is in good agreement with the $M_\mathrm{BH} = 10^{7.6\pm0.1}$~$M_\odot$ estimated from the width of its broad H$\beta$ emission in single-epoch optical spectroscopy \citep{shappee14}. For ZTF18aajupnt, we assume the H$\beta$ single-epoch spectroscopic mass of  $M_\mathrm{BH} = 10^{6.4}$~$M_\odot$ \citep{frederick19}. We note that $\lambda L_{2500\text{\normalfont\AA}}$/$L_\mathrm{Edd}$ is a UV Eddington ratio based on the 2500~\AA~luminosity rather than a bolometric Eddington ratio. The evolution of $\lambda L_{2500\text{\normalfont\AA}}$/$L_\mathrm{Edd}$ in an individual AGN outburst is predicted to show a V-shape inversion, similar to the bolometric Eddington ratio \citep[e.g., see right panel of Figure~1 in ][]{sobolewska11a}, and we compare our observations to predictions from X-ray binaries below in Section~3.1. We use this UV Eddington ratio rather than the bolometric Eddington ratio, since this avoids having to apply a bolometric correction to our observations, which can introduce additional systematic uncertainties. For example, since the SEDs of AGN may be expected to change as a function of $M_\mathrm{BH}$ (such as the theoretical $T \propto M_\mathrm{BH}^{-1/4}$ scaling of the thin accretion disk temperature at fixed Eddington ratio), it is unclear whether bolometric corrections derived from AGN with higher $M_\mathrm{BH}$ can be directly applied to AGN with lower $M_\mathrm{BH}$. Well-known bolometric corrections for AGN based on the observed optical/UV \citep[e.g.,][]{elvis94, richards06, runnoe12} and/or X-ray luminosities \citep[e.g.,][]{marconi04, vasudevan07, lusso10} are typically created using samples of AGN with black hole masses of $M_\mathrm{BH} \gtrsim$$10^8$~$M_\odot$, while the $M_\mathrm{BH}$ of our changing-look AGN here are a factor of up to $\sim$10$^{1.6}$ lower. 
	
\section{The Evolution of $\alpha_\mathrm{OX}$ \\ as a function of Eddington ratio}
\label{sec:evolution}

\subsection{Comparison of Observations to Predictions}
\label{ssc:comparison}

	The change in $\alpha_\mathrm{OX}$ as a function of Eddington ratio for each of our two changing-look AGN displays a distinct pattern. Figure~\ref{fig:vshape} (left panel) displays the observed evolution of $\alpha_\mathrm{OX}$ as a function of $\lambda L_{2500\text{\normalfont\AA}}$/$L_\mathrm{Edd}$, for both NGC~2617 and ZTF18aajupnt. ZTF18aajupnt  is at higher $\lambda L_{2500\text{\normalfont\AA}}$/$L_\mathrm{Edd}$ during these observations, and displays a positive correlation between $\alpha_\mathrm{OX}$ and $\lambda L_{2500\text{\normalfont\AA}}$/$L_\mathrm{Edd}$, such that its SED hardens ($\alpha_\mathrm{OX}$ decreases) when $\lambda L_{2500\text{\normalfont\AA}}$/$L_\mathrm{Edd}$ decreases. This correlation is well-known from single-epoch UV/X-ray observations of samples of luminous AGN, and was also observed in multi-epoch UV/X-ray observations during the fading of Mrk~1018 \citep{husemann16, noda18}. Here, we are also directly observing a SED hardening as $\lambda L_{2500\text{\normalfont\AA}}$/$L_\mathrm{Edd}$ decreases in a changing-look AGN, using well-sampled UV and X-ray light curves. In X-ray binaries, an analogous hardening of their X-ray spectra is also observed as their bolometric Eddington ratio fades towards  $L_\mathrm{bol}/L_\mathrm{Edd} \sim 10^{-2}$, and is the hallmark of an accretion state transition from the high/soft state to the low/hard state.

	In contrast to ZTF18aajupnt, NGC~2617 is at significantly lower $\lambda L_{2500\text{\normalfont\AA}}$/$L_\mathrm{Edd}$ during its outburst, and displays the opposite spectral evolution. Figure~\ref{fig:vshape} (left panel) shows for NGC~2617, $\alpha_\mathrm{OX}$ is negatively correlated with $\lambda L_{2500\text{\normalfont\AA}}$/$L_\mathrm{Edd}$, such that is SED softens ($\alpha_\mathrm{OX}$ increases) with decreasing $\lambda L_{2500\text{\normalfont\AA}}$/$L_\mathrm{Edd}$. In low-luminosity AGN, this spectral softening at low $\lambda L_{2500\text{\normalfont\AA}}$/$L_\mathrm{Edd}$ was previously only observed using single-epoch UV/X-ray observations of a sample of faded changing-look AGN, and relied on careful sample selection to avoid complications from dust extinction and a spread in black hole mass \citep{ruan19}. Here, we are directly observing this spectral softening using light curves of an individual AGN for the first time. In X-ray binaries, an analogous softening of their X-ray spectra is also observed as their bolometric luminosity fades below $\lesssim$$10^{-2}$$L_\mathrm{Edd}$, and this behavior has been suggested to be due to a change in the dominant emission mechanism \citep{sobolewska11b}.

	When combined, the observed spectral evolution of ZTF18aajupnt and NGC~2617 in Figure~\ref{fig:vshape} (left panel) traces out a V-shape inversion in the behavior of $\alpha_\mathrm{OX}$ as a function of $\lambda L_{2500\text{\normalfont\AA}}$/$L_\mathrm{Edd}$, which is predicted from X-ray binary outbursts. The observations show that the inversion occurs at a critical $\lambda L_{2500\text{\normalfont\AA}}$/$L_\mathrm{Edd} \sim 10^{-2.4}$, and the highest $\lambda L_{2500\text{\normalfont\AA}}$ data points for NGC~2617 even appear to display hints of the inversion directly. In Figure 3 (right panel), we compare these observations to predictions by \citet{sobolewska11a} for the evolution of $\alpha_\mathrm{OX}$ as a function of $\lambda L_{2500\text{\normalfont\AA}}$/$L_\mathrm{Edd}$ in an $M_\mathrm{BH} = 10^8$~$M_\odot$ AGN. These predictions are based on modeling the multi-epoch {\it Rossi X-ray Timing Explorer} (\emph{RXTE}) X-ray spectra of a prototypical outburst in the X-ray binary GRO~J1655$-$40, and then scaling the temperature of the disk component to supermassive black holes. Furthermore, the heating-to-cooling compactness ratio of the Comptonized component is assumed to be the same in AGN as inferred from GRO~J1655$-$40 (see \citealt{sobolewska11a} for details). Figure~\ref{fig:vshape} illustrates the \emph{qualitative} agreement between the observations and predictions, as both display a characteristic V-shape evolution of $\alpha_\mathrm{OX}$. However, the inversion is predicted to occur at a lower $\lambda L_{2500\text{\normalfont\AA}}$/$L_\mathrm{Edd}$ value than observed, and we explore possible reasons for this discrepancy in Section~\ref{sec:checks} of the Appendix. In Section~\ref{ssc:repro} below, we suggest a possible interpretation of the observed V-shape evolution of $\alpha_\mathrm{OX}$, based on reprocessing of X-rays.

\subsection{A Reprocessing-Based Interpretation \\ of the Inversion of $\alpha_\mathrm{OX}$}
\label{ssc:repro}

	Interpretations of our observed V-shape evolution of $\alpha_\mathrm{OX}$ as a function of $\lambda L_{2500\text{\normalfont\AA}}$/$L_\mathrm{Edd}$ in AGN as traced by changing-look AGN should take into account the time-lags detected in NGC~2617. While the right side of the V-shape evolution of $\alpha_\mathrm{OX}$ in Figure~\ref{fig:vshape} (in which the SED hardens as $\lambda L_{2500\text{\normalfont\AA}}$/$L_\mathrm{Edd}$ decreases) may be explained as an accretion state transition from the high/soft state to the low/hard state, the subsequent SED softening as $\lambda L_{2500\text{\normalfont\AA}}$/$L_\mathrm{Edd}$ decreases further is still unclear. In the data we use here, the SED softening at low Eddington ratios is probed by NGC~2617, which is well-known to display robust reverberation lags between the X-ray, UV, and optical continuum emission \citep[e.g.,][]{shappee14, oknyansky17, fausnaugh18}. \citet{shappee14} showed that these reverberation lags in the multi-wavelength light curves are well-described by a simple model where a central X-ray source above the black hole irradiates the accretion disk, which reprocesses the X-rays into UV/optical emission \citep{kazanas01, cackett07}. In this model, the UV/optical continuum during the observed outburst of NGC~2617 is dominated by this reprocessed emission, with a smaller additional contribution powered by internal dissipation in the underlying accretion disk.

	Our results suggest that the SED softening (increasing $\alpha_\mathrm{OX}$) below a critical value of $\lambda L_{2500\text{\normalfont\AA}}$/$L_\mathrm{Edd}$ in AGN may be due to reprocessing of X-ray emission. As X-ray emission from the central source is reprocessed into the UV/optical by the disk, the resultant UV/optical light curve is smoothed out in comparison to the driving X-ray light curve because of the finite light-crossing time across the reprocessing region. This smoothing is clearly observed in the X-ray and UV light curves of NGC~2617 in Figure~\ref{fig:ngc2617_lc}, where the X-ray light curve displays stronger and sharper variability in comparison to the UV light curve. Critically, this smoothing naturally causes the SED to display a softer when fainter (or conversely, harder when brighter) correlation between $\alpha_\mathrm{OX}$ and $\lambda L_{2500\text{\normalfont\AA}}$/$L_\mathrm{Edd}$. In other words, since the driving X-rays vary with greater amplitude than the UV/optical emission in Figure~\ref{fig:ngc2617_lc}, the SED hardens ($\alpha_\mathrm{OX}$ decreases) at higher $\lambda L_{2500\text{\normalfont\AA}}$/$L_\mathrm{Edd}$, and softens ($\alpha_\mathrm{OX}$ increases) at lower $\lambda L_{2500\text{\normalfont\AA}}$/$L_\mathrm{Edd}$. Thus, reprocessing of X-rays can naturally produce the left side of the V-shape evolution of $\alpha_\mathrm{OX}$ in Figure~\ref{fig:vshape}.

	We describe a qualitative picture of the full V-shape evolution of $\alpha_\mathrm{OX}$ in Figure~\ref{fig:vshape}, for an AGN fading from high $L_\mathrm{bol}/L_\mathrm{Edd} \gtrsim 10^{-1}$ to low $L_\mathrm{bol}/L_\mathrm{Edd} \lesssim 10^{-4}$. A luminous AGN in the high/soft state (i.e., at high $L_\mathrm{bol}/L_\mathrm{Edd} \gtrsim 10^{-1}$) has a soft SED (high $\alpha_\mathrm{OX}$) due to strong UV/optical emission from its luminous accretion disk. As its $L_\mathrm{bol}/L_\mathrm{Edd}$ decreases towards $L_\mathrm{bol}/L_\mathrm{Edd} \sim 10^{-2}$, the AGN undergoes an accretion state transition into the low/hard state and its SED hardens ($\alpha_\mathrm{OX}$ decreases), possibly due to truncation of the inner disk as it evaporates into an optically thin ADAF. This produces the the right side of the V-shape evolution of $\alpha_\mathrm{OX}$ in Figure~\ref{fig:vshape}. Below a critical $L_\mathrm{bol}/L_\mathrm{Edd} \lesssim 10^{-2}$, the thermal UV/optical emission from the disk becomes sufficiently dim that Comptonized X-rays from the ADAF that are reprocessed by the disk begin to dominate the UV/optical emission, over the emission that is powered by internal dissipation in the disk. Since reprocessing creates a SED softening as $L_\mathrm{bol}/L_\mathrm{Edd}$ decreases (the softer when fainter trend) as discussed above, this causes $\alpha_\mathrm{OX}$ to invert and begin increasing as $L_\mathrm{bol}/L_\mathrm{Edd}$ decreases further, producing the the left side of the V-shape evolution of $\alpha_\mathrm{OX}$ in Figure~\ref{fig:vshape}. Furthermore, since X-ray binaries also display this characteristic SED evolution, our results suggest that reprocessing is occurring in X-ray binaries as well \citep[e.g.,][]{gierlinski08}, and is responsible for the softening of their X-ray emission at low $L_\mathrm{bol}/L_\mathrm{Edd}$ in the low/hard state. However, it is unclear whether differences in the details of reprocessing in AGN and X-ray binaries could account for the discrepancy in the exact $\lambda L_{2500\text{\normalfont\AA}}$/$L_\mathrm{Edd}$ value at which the inversion in $\alpha_\mathrm{OX}$ occurs between the observations of AGN and predictions from X-ray binaries in Figure~\ref{fig:vshape}, and we defer a more quantitative study of this reprocessing picture to a future investigation.

\section{Conclusions}
\label{sec:Conclusions}

	We trace the evolution of the UV-to-X-ray spectral index $\alpha_\mathrm{OX}$ in two changing-look AGN over a wide range of Eddington ratios. Unlike previous studies that relied on single-epoch observations of samples of AGN with a wide range of Eddington ratios, here we use light curves of two individual changing-look AGN (NGC~2617 and ZTF18aajupnt) as they vary dramatically in luminosity. We show that the combination of these two changing-look AGN alone can trace out the V-shape inversion in the evolution of $\alpha_\mathrm{OX}$ as a function of Eddington ratio, which is predicted in AGN from observations of X-ray binary outbursts. Since the evolution of $\alpha_\mathrm{OX}$ probes the geometry of the disk-corona system in AGN, this suggests that the physical changes in the accretion flows of changing-look AGN are analogous to those in X-ray binary outbursts. Furthermore, we suggest a scenario in which the observed SED softening at low Eddington ratios in both AGN and X-ray binaries are due to reprocessing of Comptonized X-rays by the accretion disk, based on the reverberation time-lags between the X-ray and UV emission in NGC~2617.

	Our results emphasize the need for contemporaneous UV and X-ray light curves in interpreting quasar variability phenomena. The $\sim$few year timescales observed for changing-look AGN at both higher Eddington ratios (e.g., ZTF18aajupnt) and lower Eddington ratios (e.g., NGC~2617) suggest that the short timescales for dramatic variability in changing-look AGN can occur over a wide range of Eddington ratios. This in turn suggests that wide-field time-domain imaging surveys such as ZTF \citep{graham19b, bellm19} should discover many more outbursting AGN, and multi-wavelength follow-up (e.g., with \emph{Swift}) over the ensuing $\sim$year will be able to follow the $\alpha_\mathrm{OX}$ evolution of a number of individual AGN. This ability to map AGN variability phenomena to those observed in X-ray binary outbursts will enable possible studies of, e.g., hysteresis behavior during AGN outburst, searches for quasi-periodic oscillations during AGN accretion state transitions, and even the launching/quenching of AGN radio jets. 
	
\acknowledgments
We thank the organizers of the `Quasars in Crisis' meeting in 2019 for stimulating discussions. J.J.R, S.F.A., and M.E. are supported by Chandra Award Number GO7-18033X and GO8-19090A, issued by the  {\it Chandra} X-ray Observatory center, which is operated by the Smithsonian Astrophysical Observatory for and on behalf of the National Aeronautics Space Administration (NASA) under contract NAS8-03060. C.L.M, P.J.G., S.F.A., and J.J.R. are supported by the National Science Foundation under Grants No. AST-1715763 and AST-1715121. J.J.R. and D.H. acknowledge support from a Natural Sciences and Engineering Research Council of Canada (NSERC) Discovery Grant, a Fonds de recherche du Qu\'ebec-Nature et Technologies (FRQNT) Nouveaux Chercheurs Grant, and support from the Canadian Institute for Advanced Research (CIFAR).  J.J.R. acknowledges funding from the McGill Trottier Chair in Astrophysics and Cosmology, the McGill Space Institute, and the Dan David Foundation.

\facility{Swift}
 
\bibliographystyle{apj}

% ============================

\begin{deluxetable*}{ccccc}
\tablecaption{{\bf \emph{Swift} X-ray and UV light curve of NGC~2617} Columns include the date of the observation, the 2500~\AA~luminosity, the 2~keV luminosity, the UV-to-X-ray spectral index, and the UV Eddington ratio.}
\tablehead{  
 \colhead{Date} &   \colhead{log($\lambda L_\mathrm{2500\text{\normalfont\AA}}$)}  &  \colhead{log($\nu L_\mathrm{2keV}$)} & \colhead{$\alpha_\mathrm{OX}$}  &  \colhead{log($\lambda L_\mathrm{2500\text{\normalfont\AA}}/L_\mathrm{Edd}$)} \\ 
 \colhead{(MJD)} &   \colhead{[erg s$^{-1}$]}  &  \colhead{[erg s$^{-1}$]}  & &  
 }
\startdata
 \vspace{2pt}
55516.8 & 42.99 $\pm$ 0.02 & 42.55 $\pm$ 0.06 & 1.17 $\pm$ 0.02 & -2.63 $\pm$ 0.06 \\
55517.5 & 42.98 $\pm$ 0.02 & 42.48 $\pm$ 0.04 & 1.19 $\pm$ 0.02 & -2.65 $\pm$ 0.04 \\
55518.6 & 43.00 $\pm$ 0.02 & 42.45 $\pm$ 0.04 & 1.21 $\pm$ 0.02 & -2.63 $\pm$ 0.04 \\
55519.5 & 43.01 $\pm$ 0.02 & 42.50 $\pm$ 0.04 & 1.20 $\pm$ 0.02 & -2.61 $\pm$ 0.04 \\
55521.0 & 43.10 $\pm$ 0.02 & 42.70 $\pm$ 0.05 & 1.15 $\pm$ 0.02 & -2.53 $\pm$ 0.05 \\
55521.8 & 43.15 $\pm$ 0.01 & 42.83 $\pm$ 0.05 & 1.12 $\pm$ 0.02 & -2.47 $\pm$ 0.05 \\
55523.4 & 43.14 $\pm$ 0.01 & 43.03 $\pm$ 0.04 & 1.04 $\pm$ 0.02 & -2.49 $\pm$ 0.04 \\
55524.2 & 43.21 $\pm$ 0.01 & 43.08 $\pm$ 0.04 & 1.05 $\pm$ 0.02 & -2.41 $\pm$ 0.04 \\
55524.8 & 43.19 $\pm$ 0.02 & 43.09 $\pm$ 0.04 & 1.04 $\pm$ 0.02 & -2.43 $\pm$ 0.04 \\
55526.3 & 43.22 $\pm$ 0.01 & 43.05 $\pm$ 0.04 & 1.07 $\pm$ 0.02 & -2.40 $\pm$ 0.05 \\
55528.0 & 43.22 $\pm$ 0.01 & 42.97 $\pm$ 0.04 & 1.10 $\pm$ 0.02 & -2.40 $\pm$ 0.04 \\
55531.6 & 43.23 $\pm$ 0.02 & 42.94 $\pm$ 0.06 & 1.11 $\pm$ 0.02 & -2.40 $\pm$ 0.06 \\
55533.1 & 43.28 $\pm$ 0.02 & 42.94 $\pm$ 0.05 & 1.13 $\pm$ 0.02 & -2.34 $\pm$ 0.05 \\
55534.3 & 43.22 $\pm$ 0.02 & 42.96 $\pm$ 0.05 & 1.10 $\pm$ 0.02 & -2.40 $\pm$ 0.05 \\
55535.2 & 43.32 $\pm$ 0.02 & 42.97 $\pm$ 0.04 & 1.13 $\pm$ 0.02 & -2.30 $\pm$ 0.04 \\
55536.1 & 43.28 $\pm$ 0.01 & 43.00 $\pm$ 0.04 & 1.11 $\pm$ 0.02 & -2.35 $\pm$ 0.04 \\
55536.9 & 43.26 $\pm$ 0.01 & 43.02 $\pm$ 0.04 & 1.09 $\pm$ 0.02 & -2.36 $\pm$ 0.04 \\
55537.9 & 43.27 $\pm$ 0.01 & 43.04 $\pm$ 0.04 & 1.09 $\pm$ 0.02 & -2.35 $\pm$ 0.04 \\
55539.3 & 43.24 $\pm$ 0.01 & 43.05 $\pm$ 0.04 & 1.07 $\pm$ 0.01 & -2.39 $\pm$ 0.04 \\
55540.1 & 43.21 $\pm$ 0.01 & 43.04 $\pm$ 0.04 & 1.06 $\pm$ 0.02 & -2.41 $\pm$ 0.04 \\
55541.1 & 43.21 $\pm$ 0.02 & 43.03 $\pm$ 0.04 & 1.07 $\pm$ 0.02 & -2.42 $\pm$ 0.04 \\
55542.0 & 43.21 $\pm$ 0.01 & 43.02 $\pm$ 0.04 & 1.07 $\pm$ 0.02 & -2.41 $\pm$ 0.04 \\
55543.4 & 43.18 $\pm$ 0.01 & 42.99 $\pm$ 0.04 & 1.07 $\pm$ 0.02 & -2.44 $\pm$ 0.04 \\
55544.3 & 43.15 $\pm$ 0.02 & 42.94 $\pm$ 0.04 & 1.08 $\pm$ 0.02 & -2.47 $\pm$ 0.04 \\
55545.3 & 43.12 $\pm$ 0.02 & 42.84 $\pm$ 0.04 & 1.11 $\pm$ 0.02 & -2.50 $\pm$ 0.04 \\
55546.0 & 43.07 $\pm$ 0.02 & 42.74 $\pm$ 0.04 & 1.13 $\pm$ 0.02 & -2.56 $\pm$ 0.04 \\
55546.5 & 43.06 $\pm$ 0.02 & 42.65 $\pm$ 0.04 & 1.16 $\pm$ 0.02 & -2.56 $\pm$ 0.04 \\
55547.5 & 43.05 $\pm$ 0.02 & 42.47 $\pm$ 0.04 & 1.22 $\pm$ 0.02 & -2.57 $\pm$ 0.04 \\
55549.3 & 43.03 $\pm$ 0.02 & 42.25 $\pm$ 0.05 & 1.30 $\pm$ 0.02 & -2.60 $\pm$ 0.05 \\
55550.7 & 43.01 $\pm$ 0.02 & 42.30 $\pm$ 0.05 & 1.27 $\pm$ 0.02 & -2.61 $\pm$ 0.05 \\
55551.6 & 43.03 $\pm$ 0.02 & 42.41 $\pm$ 0.05 & 1.24 $\pm$ 0.02 & -2.60 $\pm$ 0.05 \\
55554.6 & 43.01 $\pm$ 0.02 & 42.81 $\pm$ 0.05 & 1.08 $\pm$ 0.02 & -2.61 $\pm$ 0.05 \\
55558.5 & 42.91 $\pm$ 0.02 & 42.21 $\pm$ 0.08 & 1.27 $\pm$ 0.03 & -2.71 $\pm$ 0.08 \\
55558.8 & 42.90 $\pm$ 0.02 & 42.15 $\pm$ 0.08 & 1.29 $\pm$ 0.03 & -2.73 $\pm$ 0.08 \\
55559.6 & 42.91 $\pm$ 0.02 & 42.07 $\pm$ 0.07 & 1.32 $\pm$ 0.03 & -2.71 $\pm$ 0.07 \\
55560.8 & 42.92 $\pm$ 0.02 & 42.06 $\pm$ 0.06 & 1.33 $\pm$ 0.02 & -2.70 $\pm$ 0.06 \\
55562.9 & 42.99 $\pm$ 0.02 & 42.39 $\pm$ 0.05 & 1.23 $\pm$ 0.02 & -2.63 $\pm$ 0.05 \\
55563.6 & 42.97 $\pm$ 0.02 & 42.53 $\pm$ 0.08 & 1.17 $\pm$ 0.03 & -2.66 $\pm$ 0.08 \\
\enddata
\label{tab:tab1}
\end{deluxetable*}
% ============================

% ============================

\begin{deluxetable*}{cccccc}
\tablecaption{{\bf \emph{Swift} X-ray and UV light curves of ZTF18aajupnt} Columns include the date of the observation, the 2500~\AA~luminosity, the 2~keV luminosity, the UV-to-X-ray spectral index, and the UV Eddington ratio.
}
\tablehead{  
 \colhead{Date} &   \colhead{log($\lambda L_\mathrm{2500\text{\normalfont\AA}}$)}  &  \colhead{log($\nu L_\mathrm{2keV}$)} & \colhead{$\alpha_\mathrm{OX}$}  &  \colhead{log($\lambda L_\mathrm{2500\text{\normalfont\AA}}/L_\mathrm{Edd}$)} \\ 
 \colhead{(MJD)} &   \colhead{[erg s$^{-1}$]}  &  \colhead{[erg s$^{-1}$]}  & &  
 }
\startdata
 \vspace{2pt}
58329 & 42.75 & 40.86 & 1.73 & -1.76 \\
58342 & 42.66 & 41.20 & 1.56 & -1.85 \\
58350 & 42.59 & 41.42 & 1.45 & -1.93 \\
58352 & 42.60 & 41.18 & 1.54 & -1.92 \\
58357 & 42.62 & 41.40 & 1.47 & -1.89 \\
58362 & 42.59 & 41.59 & 1.38 & -1.93 \\
58379 & 42.56 & 41.51 & 1.40 & -1.95 \\
58384 & 42.59 & 41.68 & 1.35 & -1.93 \\
58389 & 42.51 & 41.59 & 1.36 & -2.00 \\
58394 & 42.58 & 41.80 & 1.30 & -1.94 \\
58399 & 42.51 & 41.76 & 1.29 & -2.00 \\
58404 & 42.49 & 41.80 & 1.26 & -2.03 \\
58445 & 42.44 & 41.71 & 1.28 & -2.07 \\
58450 & 42.42 & 41.64 & 1.30 & -2.09 \\
58455 & 42.47 & 41.85 & 1.24 & -2.04 \\
58460 & 42.44 & 41.87 & 1.22 & -2.07 \\
58559 & 42.31 & 41.86 & 1.17 & -2.20 \\
\enddata
\label{tab:tab2}
\end{deluxetable*}
% ============================
% ============================

\begin{appendix}

\section{Consistency Checks}
\label{sec:checks}
	Although our observations of changing-look AGN in Figure~\ref{fig:vshape} (left panel) are in broad agreement with the predictions from X-ray binary outbursts in Figure~\ref{fig:vshape} (right panel), the V-shape inversion is observed to occur at a critical UV Eddington ratio of $\lambda L_{2500\text{\normalfont\AA}}$/$L_\mathrm{Edd} \sim 10^{-2.4}$, while the predictions suggest that it should instead occur at a lower value of $\lambda L_{2500\text{\normalfont\AA}}$/$L_\mathrm{Edd} \sim 10^{-3}$. We discuss three possible reasons for this $\sim$0.6~dex discrepancy, including the effects of differences in black hole mass, possible issues in our subtraction of the host galaxy starlight, and challenges in generating the predictions from observations of X-ray binary outbursts. It is unclear if any of these possibilities can account for the discrepancy.

% ============================

\subsection{Effects of Differences in Black Hole Mass}
\label{ssc:bhmass}

	Differences in the black hole masses of the observations and predictions can lead to systematic differences in $\lambda L_{2500\text{\normalfont\AA}}$/$L_\mathrm{Edd}$. In Figure~\ref{fig:vshape}, the predictions assume a black hole mass of $M_\mathrm{BH} = 10^8$~$M_\odot$, while our changing-look AGN have lower masses ($10^{7.5}$~$M_\odot$ for NGC~2617, and $10^{6.4}$~$M_\odot$ for ZTF18aajupnt). Since the SED shapes of black hole accretion flows are expected to change as a function of $M_\mathrm{BH}$, $\lambda L_{2500\text{\normalfont\AA}}$/$L_\mathrm{Edd}$ can be systematically different for our observed changing-look AGN in comparison to the predictions that assume $M_\mathrm{BH} = 10^8$~$M_\odot$. For example, since the thin disk emission in AGN with lower $M_\mathrm{BH}$ is expected to peak at higher frequencies, $\lambda L_{2500\text{\normalfont\AA}}$/$L_\mathrm{Edd}$ for AGN at different black hole masses will not probe the same part of the SED. To gauge whether this effect strongly affects our results, we compare the expected difference in $\lambda L_{2500\text{\normalfont\AA}}$/$L_\mathrm{Edd}$ for accretion disks with mass $M_\mathrm{BH} = 10^8$~$M_\odot$ (assumed in the predictions in Figure~\ref{fig:vshape}) and $10^{7.5}$~$M_\odot$ (the mass of NGC~2617, which probes the inversion in Figure~\ref{fig:vshape}). We compute theoretical optical/UV SEDs of a standard Shakura-Sunyaev thin accretion disk \citep{shakura73}, assuming an Eddington ratio of $10^{-2}$ and an inner edge at 6$r_\mathrm{g}$ (where $r_\mathrm{g}$ is the gravitational radius). We find that $\lambda L_{2500\text{\normalfont\AA}}$/$L_\mathrm{Edd}$ is 0.20~dex \emph{lower} for an AGN with $M_\mathrm{BH} = 10^{7.5}$~$M_\odot$ than an AGN with $M_\mathrm{BH} = 10^8$~$M_\odot$. We also empirically test this conclusion using the reverberation-mapped sample of 29 AGN from \citet{vasuvedan09}, who measured $M_\mathrm{BH}$, $\Gamma$, $\nu L_\mathrm{2-10leV}$, $L_\mathrm{bol}/L_\mathrm{Edd}$, and $\alpha_\mathrm{OX}$ using contemporaneous UV/optical and X-ray observations from \emph{XMM-Newton}. Similar to our theoretical test, we select a sub-sample of objects (Fairall~9, Mrk~590, Mrk~79, NGC~3516, NGC~5548, PG~2130+099) that lie within both a narrow range of low Eddington ratios of $10^{-2.5} < L_\mathrm{bol}/L_\mathrm{Edd} < 10^{-1.5}$, and black hole masses of $10^7 < M_\mathrm{BH} < 10^9$. We divide these 6 objects into a high-$M_\mathrm{BH}$ subsample with mean $\langle$$M_\mathrm{BH}$$\rangle$$ =10^{8.3}$ $M_\odot$, and a low-$M_\mathrm{BH}$ subsample with mean $\langle$$M_\mathrm{BH}$$\rangle$$ =10^{7.7}$ $M_\odot$. We find that $\lambda L_{2500\text{\normalfont\AA}}$/$L_\mathrm{Edd}$ is 0.30~dex \emph{lower} for the  low-$M_\mathrm{BH}$ subsample than for the high-$M_\mathrm{BH}$ subsample. Thus, our theoretical and empirical tests suggest that that differences in $M_\mathrm{BH}$ are unlikely to be the origin of the discrepancy in the $\lambda L_{2500\text{\normalfont\AA}}$/$L_\mathrm{Edd}$ at which the inversion occurs between the observations and predictions, because carefully accounting for this effect would actually sightly \emph{increase} the discrepancy. 

% ============================

\subsection{Host Galaxy Subtraction}
\label{ssc:host}

\begin{figure*} [t!]
\center{
\includegraphics[scale=0.62,angle=0]{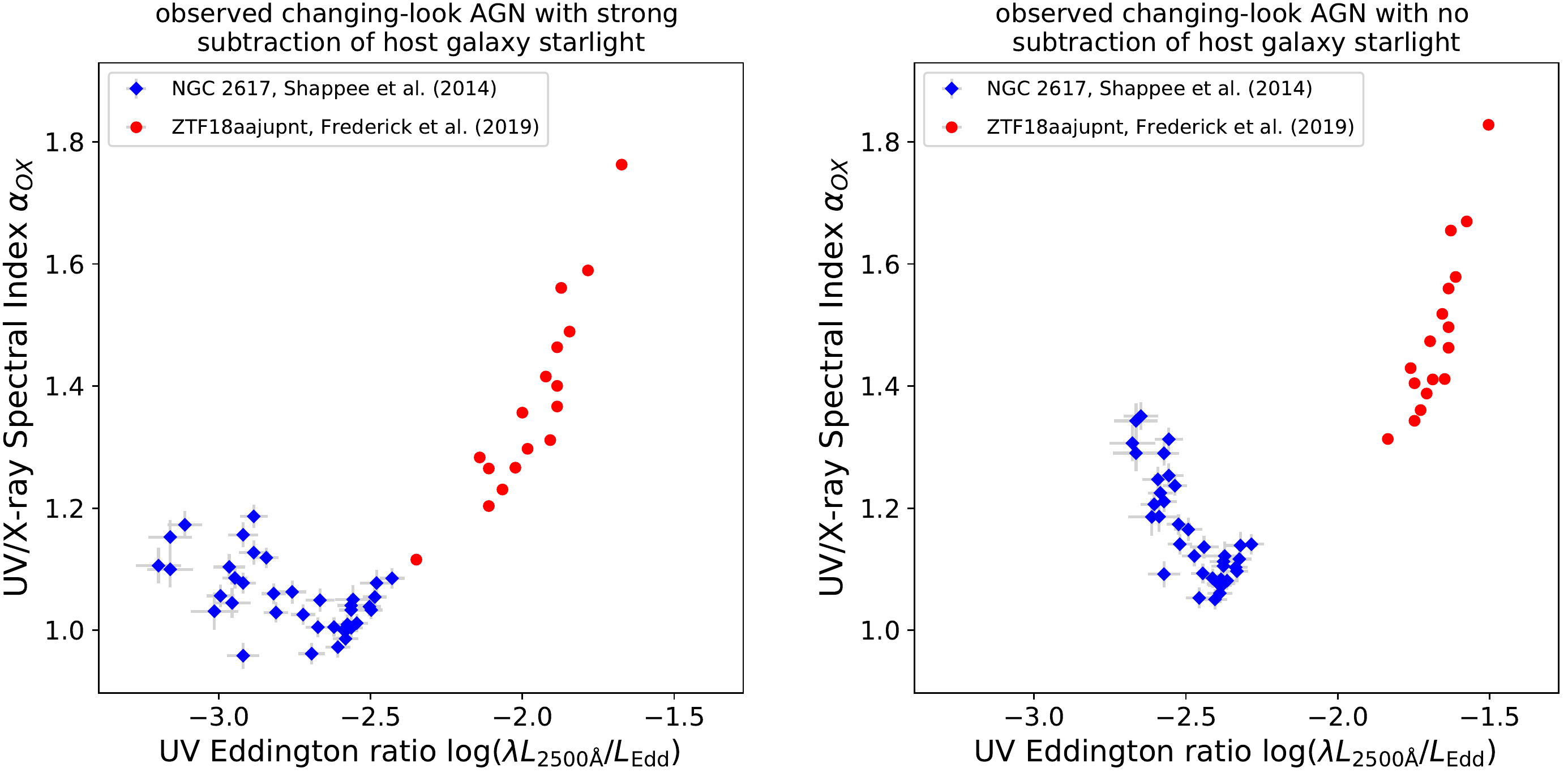}
}
\figcaption{Our results on the V-shape inversion in the $\alpha_\mathrm{OX}$ evolution of changing-look AGN are not strongly dependent on our subtraction of the host galaxy starlight. \emph{Left}: The evolution of the UV-to-X-ray spectral index $\alpha_\mathrm{OX}$ as a function of the UV Eddington ratio $\lambda L_{2500\text{\normalfont\AA}}$/$L_\mathrm{Edd}$ (similar to Figure~\ref{fig:vshape}), after subtracting a stronger host galaxy component from  $\lambda L_{2500\text{\normalfont\AA}}$ (see Section~\ref{ssc:host}). \emph{Right}:  The evolution of the UV-to-X-ray spectral index $\alpha_\mathrm{OX}$ as a function of the UV Eddington ratio $\lambda L_{2500\text{\normalfont\AA}}$/$L_\mathrm{Edd}$ (similar to the left panel), but without subtracting a host galaxy component from  $\lambda L_{2500\text{\normalfont\AA}}$ (see Section~\ref{ssc:host}). In either case of assuming a strong host galaxy component or no host galaxy component, the V-shape inversion does not shift to significantly different values of $\lambda L_{2500\text{\normalfont\AA}}$/$L_\mathrm{Edd}$. Thus, the details of our host galaxy subtraction are unlikely to be the origin of the discrepancy between the observations and predictions in Figure~\ref{fig:vshape}. 
}
\label{fig:vshape_nohostgalaxysub}
\end{figure*}

	Differences in the host galaxy subtraction procedure can also lead to divergent values of $\lambda L_{2500\text{\normalfont\AA}}$/$L_\mathrm{Edd}$. Since we subtracted an estimate of the host galaxy contribution from the $\lambda L_{2500\text{\normalfont\AA}}$ luminosities, it is possible that the observed inversion at $\lambda L_{2500\text{\normalfont\AA}}$/$L_\mathrm{Edd} \sim 10^{-2.4}$ may shift to a different value of $\lambda L_{2500\text{\normalfont\AA}}$/$L_\mathrm{Edd}$ if a different host galaxy luminosity is assumed. We test this empirically, by changing our assumed host galaxy luminosities to higher and lower values that likely bracket the true value, and re-creating Figure~\ref{fig:vshape} to display the results. We first assume that the host galaxy contributions to $\lambda L_{2500\text{\normalfont\AA}}$ for both NGC~2617 and ZTF18aajupnt are 80\% of the faintest data point in their respective $\lambda L_{2500\text{\normalfont\AA}}$ light curves. In this strong host galaxy contribution scenario, the host galaxy $\lambda L_{2500\text{\normalfont\AA}}$ for NGC~2617 and ZTF18aajupnt are 10$^{42.8}$~erg~s$^{-1}$ and 10$^{42.54}$~erg~s$^{-1}$, respectively. This is in contrast to the fiducial host galaxy $\lambda L_{2500\text{\normalfont\AA}}$ of 10$^{42.0}$~erg~s$^{-1}$ and 10$^{42.44}$~erg~s$^{-1}$ we previously assumed for NGC~2617 and ZTF18aajupnt respectively in Figure~\ref{fig:vshape}. The resulting $\alpha_\mathrm{OX}$ evolution as a function of $\lambda L_{2500\text{\normalfont\AA}}$/$L_\mathrm{Edd}$ for this strong host galaxy subtraction case is shown in Figure~\ref{fig:vshape_nohostgalaxysub} (left panel). We then assume low host galaxy contributions to $\lambda L_{2500\text{\normalfont\AA}}$ of 0 for both NGC~2617 and ZTF18aajupnt (i.e., we do not perform any host galaxy subtraction), and show the resulting $\alpha_\mathrm{OX}$ evolution for this no host galaxy subtraction case in Figure~\ref{fig:vshape_nohostgalaxysub} (right panel). A comparison of Figure~\ref{fig:vshape_nohostgalaxysub} to our fiducial results in Figure~\ref{fig:vshape} demonstrates that (a) the observed V-shaped inversion in the evolution of $\alpha_\mathrm{OX}$ is not strongly dependent on the host galaxy subtraction, and (b) the discrepancy between the observations and predictions for the critical $\lambda L_{2500\text{\normalfont\AA}}$/$L_\mathrm{Edd}$ value at which the $\alpha_\mathrm{OX}$ inversion occurs persists even when changing the assumed host galaxy luminosity over a wide range of values. This insensitivity of the inflection point in Figure~\ref{fig:vshape_nohostgalaxysub} to the host galaxy subtraction can be understood by considering how the data points in Figure~\ref{fig:vshape_nohostgalaxysub} move when different host galaxy luminosities are assumed. Subtracting a host galaxy contribution from the observed $\lambda L_{2500\text{\normalfont\AA}}$ light curve of an individual object will cause the $\lambda L_{2500\text{\normalfont\AA}}$/$L_\mathrm{Edd}$ values to decrease more strongly for data points at low $\lambda L_{2500\text{\normalfont\AA}}$ than for data points at high $\lambda L_{2500\text{\normalfont\AA}}$. This effect can be seen in Figure~\ref{fig:vshape_nohostgalaxysub}, where the low-$\lambda L_{2500\text{\normalfont\AA}}$/$L_\mathrm{Edd}$ data points for both NGC~2617 and ZTF18aajupnt shift to the left significantly more when subtracting a strong host galaxy contribution, while the high $\lambda L_{2500\text{\normalfont\AA}}$/$L_\mathrm{Edd}$ data points do not shift as strongly. Since the inflection point in Figure~\ref{fig:vshape_nohostgalaxysub} is probed primarily by the high-$\lambda L_{2500\text{\normalfont\AA}}$/$L_\mathrm{Edd}$ data points of NGC~2617, this causes the inflection point to be rather insensitive to the host galaxy subtraction.
	
% ============================

\subsection{Systematics in Predictions from X-ray Binaries}
\label{ssc:systematics}

	Lastly, the X-ray binary spectral modeling that produced the predictions in Figure~\ref{fig:vshape} (right panel) may systematically under-predict the $\lambda L_{2500\text{\normalfont\AA}}$/$L_\mathrm{Edd}$ value at which the inflection point occurs. These predictions are based on modeling the \emph{RXTE} multi-epoch X-ray spectra throughout the outburst of GRO~J1655$-$40 using thin disk and Comptonized coronal spectral components. At lower Eddington ratios during the outburst, the accretion disk emission shifts to softer X-rays due to a decrease in the effective disk temperature. In combination with \emph{RXTE}'s relatively low sensitivity to soft X-rays, measuring the thin disk temperature is more difficult in this regime. Any systematic errors in the measured thin disk temperature at lower Eddington ratios will then propagate into the predictions for the $\alpha_\mathrm{OX}$ evolution of AGN, especially at the lower $\lambda L_{2500\text{\normalfont\AA}}$/$L_\mathrm{Edd}$ where the V-shape inversion occurs. We suggest two possible future approaches to test this possibility. From the X-ray binary side, additional X-ray spectral modeling of X-ray binary outbursts (especially as they fade to low Eddington ratios) and scaling to supermassive black holes can test whether the predicted inversion in $\alpha_\mathrm{OX}$ indeed occurs at $\lambda L_{2500\text{\normalfont\AA}}$/$L_\mathrm{Edd} \sim 10^{-3}$ as inferred from GRO~J1655$-$40. For example, future X-ray observations that follow the fading of the recent outbursts of MAXI~J1820+070 \citep[e.g.,][]{kawamuro18, tucker18} into quiescence can improve predictions for the inversion of $\alpha_\mathrm{OX}$. This may be performed using the \emph{Neutron Star Interior Composition Explorer} \citep[\emph{NICER;}][]{gendreau16}, thanks to its higher sensitivity to soft X-rays from the faint thin disk spectral component at low Eddington ratios. From the AGN side, additional UV/X-ray observations of changing-look AGN can further test whether the inversion of $\alpha_\mathrm{OX}$ is indeed observed to occur at  $\lambda L_{2500\text{\normalfont\AA}}$/$L_\mathrm{Edd} \sim 10^{-2.4}$. This may soon be possible with additional \emph{Swift} monitoring of ZTF18aajupnt, which might reveal fading in the X-rays, and an inversion in its $\alpha_\mathrm{OX}$ evolution. 

 %------- FIGURE 4 -------
% ============================

\end{appendix}

\end{document}